\documentclass[]{article}

\usepackage[T1]{fontenc}
\usepackage[left=1.35in,top=1in,right=1.35in,bottom=1in]{geometry}
\usepackage[latin9]{inputenc}
\usepackage{graphicx}
\usepackage{booktabs}
\usepackage[figuresright]{rotating}
\usepackage{pdflscape}
\usepackage{tabularx}
\usepackage[super]{nth}
\usepackage{amssymb}
\usepackage{csquotes} 
\usepackage[inline]{enumitem}
\usepackage{scrextend}
\usepackage{epstopdf}
\usepackage{float}
\usepackage{natbib}[8.31] 
\usepackage[table]{xcolor}
\usepackage{url}
\usepackage{amsmath}
\usepackage{adjustbox}

\begin{document}


\title{Quasi-random Monte Carlo application in CGE systematic sensitivity analysis}

\author{Theodoros Chatzivasileiadis \\ Institute for Environmental Studies, Vrije Universiteit Amsterdam, The Netherlands\thanks{Contact T. Chatzivasileiadis, Email: t.chatzivasileiadis@vu.nl, Address: De Boelelaan 1087, 1081 HV Amsterdam, The Netherlands.}}

\date{}
\maketitle

\begin{abstract}
The uncertainty and robustness of Computable General Equilibrium models can be assessed by conducting a Systematic Sensitivity Analysis. Different methods have been used in the literature for SSA of CGE models such as Gaussian Quadrature and Monte Carlo methods. This paper explores the use of Quasi-random Monte Carlo methods based on the Halton and Sobol' sequences as means to improve the efficiency over regular Monte Carlo SSA, thus reducing the computational requirements of the SSA. The findings suggest that by using low-discrepancy sequences, the number of simulations required by the regular MC SSA methods can be notably reduced, hence lowering the computational time required for SSA of CGE models.  
\end{abstract}



\section{Introduction}
The use of Systematic Sensitivity Analysis (SSA) in the context of Computable General Equilibrium (CGE) models is gaining a new momentum within the CGE literature. SSA sought to address the input and calibration-data uncertainty of the models in order to capture the uncertainties surrounding CGE models and their results. This paper discusses the use of quasi-random Monte Carlo methods in SSA of CGE models.

We begin by defining the general form of a computable general equilibrium model as:
\begin{equation}\label{eq:cge}
G(x,\beta)=0
\end{equation} 

The most common approach of SSA in CGE models is based on Gaussian Quadrature (GQ) design. That choice is based on the fact that GQ requires only a few data points to approximate the central moments of stochastic variables, making this method very computational inexpensive. An alternative to the GQ is the regular Monte Carlo (rMC) SSA that requires a large amount of realisations such that: the mean ($\bar{x}$) in the univariate case as defined by equation~\ref{eq:mean}, is an unbiased estimator of $K(\beta)$,

\begin{equation}\label{eq:mean}
I=E[K(\beta)]=\int_a^b K(\beta)p(\beta)d\beta =\bar{x} 
\end{equation}

\begin{equation}\label{eq:mean2}
I_{N}=\sum_{n=1}^{N}w_{n}K(\beta_{n})
\end{equation}

where $w_{n}=\frac{1}{N}$ for each realization n, x (eq. \ref{eq:cge}) represents a vector of results or endogenous variables (such as prices, welfare etc.), $\beta$ (eq. \ref{eq:cge}) is a vector of exogenous variables, $ x^{\star}(\beta) \equiv K(\beta)$ is a vector of results for each given parameter $\beta$ and $p$ is the non-zero probability density function (pdf).

The difference between QG and rMC is in the way $w_{n}$ is defined. While in rMC all realizations have equal weight 1/N, in QG we choose the most appropriate points within the interval [a, b] and associated weights $w_{n}$, such that, the crude moments of the approximating distribution equals the moments of the true distribution from zero to some specified order \cite{TC2}. Thus the GQ method is able to economise the computational requirements of the SSA (i.e small number of simulations are required).

Economising the SSA through GQ is not without its drawbacks. For example, a significant amount of information is lost regarding the shape of the distribution, its higher-order moments, and its range. This information might be important when the shocks applied to the CGE model are expected to have asymmetric impacts\footnote{Additional information about the differences between GQ and rMC can be found in \cite{TC2} and \cite{VP}.}. 

By using rMC SSA instead, we sample from the total distribution of inputs, thus avoiding the GQ drawbacks at the cost of computation time. Moreover, the number of points required by the GQ is based on the shocks evaluated and the degree of the quadratures. For example in the case of \cite{TC2}, 4096 simulations were required by the Liu quadrature based on the GEMPACK software. Thus, in that case, the advantage of GQ that economises the SSA is questioned. \textit{'{[...]An advantage of the rMC method, compared to GQ, lies on the estimation of the error. In the rMC method, the error is estimated from the generated data, whereas in the QG more global measures of error estimation are required such as the Chebyshev's inequality for the confidence bounds. The Chebyshev's inequality, will produce confidence bounds that are extremely conservative compared to the Central Limit Theorem which provides narrower confidence intervals if the available number of data points is sufficiently large}{[...]}'}\cite{TC2,VP}. 

From the above, the conundrum choosing between; economising the SSA with GQ and gaining the advantages of rMC SSA is apparent. This paper discusses a way to keep the advantages of the rMC SSA methodology while minimising the number of simulations required. We contribute to the existing literature by applying a Quasi-Monte Carlo (QMC) SSA on a static CGE model in order to economise the SSA without loosing important information as in the GQ SSA, in a computationally inexpensive way, in contrast to the rMC SSA.

\section{Methods and data}
\subsection{rMC and QMC}
Based on the model of \cite{TC1} and the input SSA discussed in \cite{TC2}, we explore the use of quasi-random realisations in the MC SSA of CGE models. We begin with a brief description of the differences between rMC and QMC SSA.

Going back to equation \ref{eq:mean}, according to the Strong Law of Large Numbers for the rMC method, the approximation is convergent with probability one, i.e:

\begin{equation}
\lim_{N\to\infty} I_{N} \rightarrow I
\end{equation}

Then the error of the MC integration is:
\begin{equation}
\epsilon_{N}=I-I_{N} \approx \sigma N^{-\frac{1}{2}} x
\end{equation}
if N is sufficiently large based on the Central Limit Theorem, where $\sigma$ is 
\begin{equation}\label{eq:var}
\sqrt{Var(x)}=\bigg[\int_a^b \bigg(K(\beta)-E[K(\beta)]\bigg)^{2}p(\beta)\mathrm{d}\beta\bigg]^{\frac{1}{2}}
\end{equation}

Based on the above, the convergence rate of the rMC is $O(N^{-\frac{1}{2}})$ which is independent of dimensions.  

In the case of rMC, we randomly select points and proximate equation \ref{eq:mean} by the empirical average in equation \ref{eq:mean2}. In QMC the points are selected semi-deterministically, such that the chosen points provide the best possible spread. The chosen points are highly equidistributed thus providing greater uniformity compared to the pseudo-random numbers. In QMC, the resulting convergence rate is $O((\log N)^{k} N^{-1})$, which is dependent on the k dimensions\footnote{See \cite{caflisch} for more information.}. As a result, for the same number of evaluations, the QMC method, using low-discrepancy sequences, achieves higher accuracy thus faster convergence. This characteristic of the QMC method is very appealing in our case. We are interested in faster convergence, thus smaller number of simulations required to conduct a SSA of our CGE model.

One issue with QMC is that since the convergence is dependent on the dimensions of the problem, improved accuracy is lost in problems of high dimension. Another problem is due to existing correlations between the points of the quasi-random sequence used. The latter problem can be resolved by skip, leap over, or scramble the values in the sequence.

\subsection{SSA input generation}
In our analysis we have used two different low-discrepancy sequences: the Halton; and Sobol' sequences \cite{jank}. The Halton sequence is generated by using different prime bases to generate the sequence, where for the Sobol' a base of 2 is used with a reordering of the coordinates in each dimension. In the Halton sequence we skip 1,000 and leap 100 using \textit{RR2} scrambler. The Sobol' sequence is generated using leap 10,000 and leap 100 using the \textit{MatousekAffineOwen} scrambler.

Based on \cite{TC2}, we have 13 regions and a total of 39 shocks applied to the CGE model. The process initially used for the rMC is the following:

\begin{enumerate}
	\item \textit{Generate 39 pseudo-random samples (10,000 realisations) based on the uniform distribution;}
	\item \textit{Transform to the triangular distribution given the minimum, median and maximum for each regional changes of Land, Capital and Productivity;}
	\item \textit{Run 10,000 simulations based on these inputs}.
\end{enumerate}

In the QMC method the only difference is in step 1. Instead of the 39 pseudo-random samples we generate 39 quasi-random sequences based on the Halton and Sobol' sequences. Steps 2 and 3 are followed as described above.

\section{Results}
For simplicity, we focus on one result of the CGE model; the Hicksian Equivalent Variation (HEV) for two of the regions only: Central-Asia and North-America. The choice of the regional results is arbitrary since all regions show similar differences between the rMC and the two QMC SSA methods. 

\begin{figure}[H]
	\centering
	\includegraphics[scale=0.45]{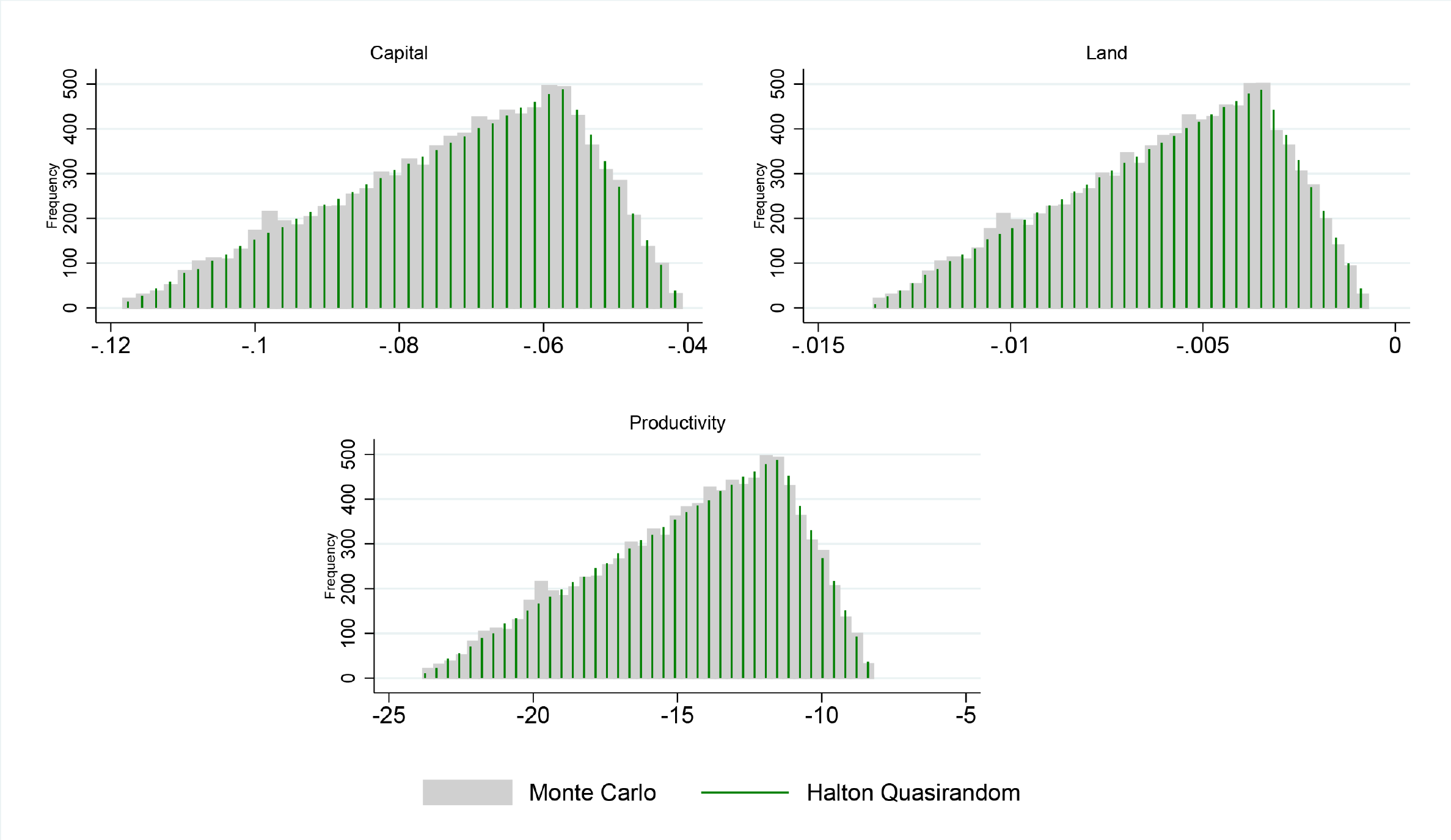}
	\caption{Shocks Histogram based on the triangular distribution using pseudo-random sampling and Halton sequence for Central Asia}
	\label{fig:CAH}
\end{figure}

\begin{figure}[H]
	\centering
	\includegraphics[scale=0.45]{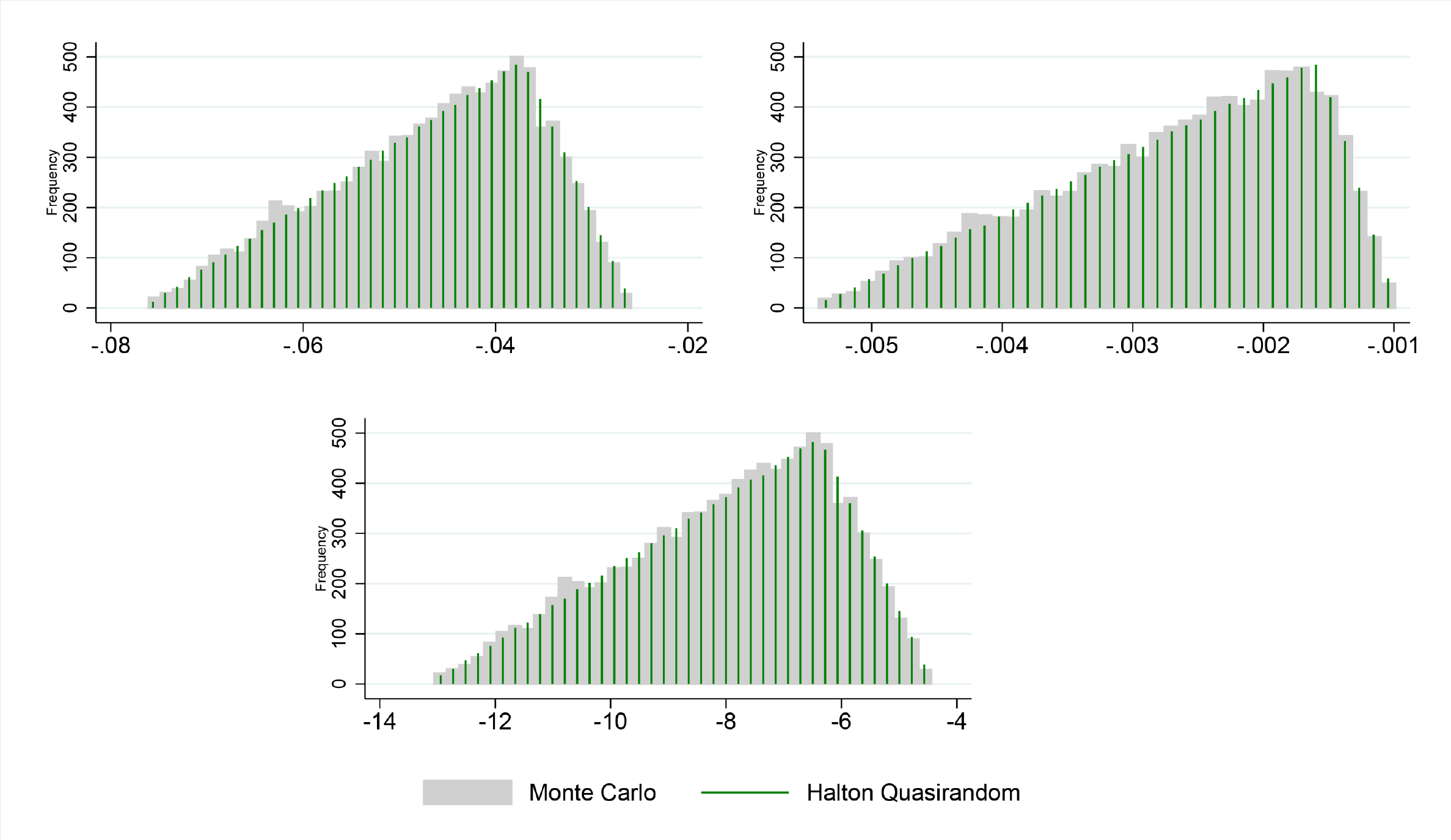}
	\caption{Shocks Histogram based on the triangular distribution using pseudo-random sampling and Halton sequence for North America}
	\label{fig:NAH}
\end{figure}

First we discuss the differences between the shocks' distributions. Here we only present the Halton sequence-based inputs, since we saw no substantial difference in the histogram with the Sobol' generated inputs. Looking at both figures \ref{fig:CAH} and \ref{fig:NAH} for Land, Capital and Productivity shocks, we see the expected; the QMC shocks are evenly distributed without any clustering (i.e spikes in the histogram as around -20 for Central Asia's productivity in figure \ref{fig:CAH}) compared to the rMC shocks' histograms.

\begin{figure}[H]
	\centering
	\includegraphics[scale=0.5]{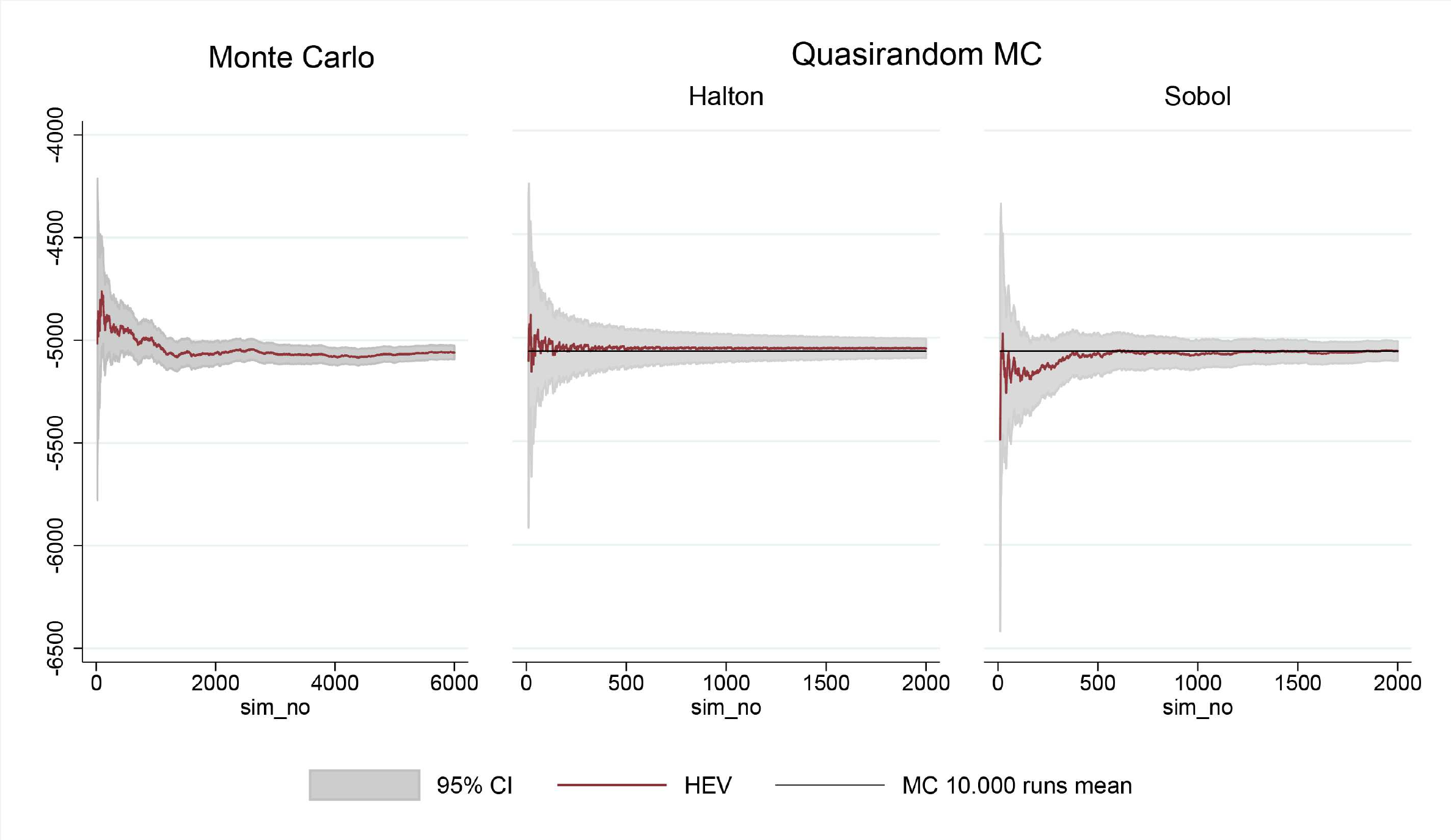}
	\caption{DIVA-C HEV mean convergence for the triangular distribution using: random sampling; Halton; and Sobol sequences for Central Asia}
	\label{fig:CAM}
\end{figure}

\begin{figure}[H]
	\centering
	\includegraphics[scale=0.5]{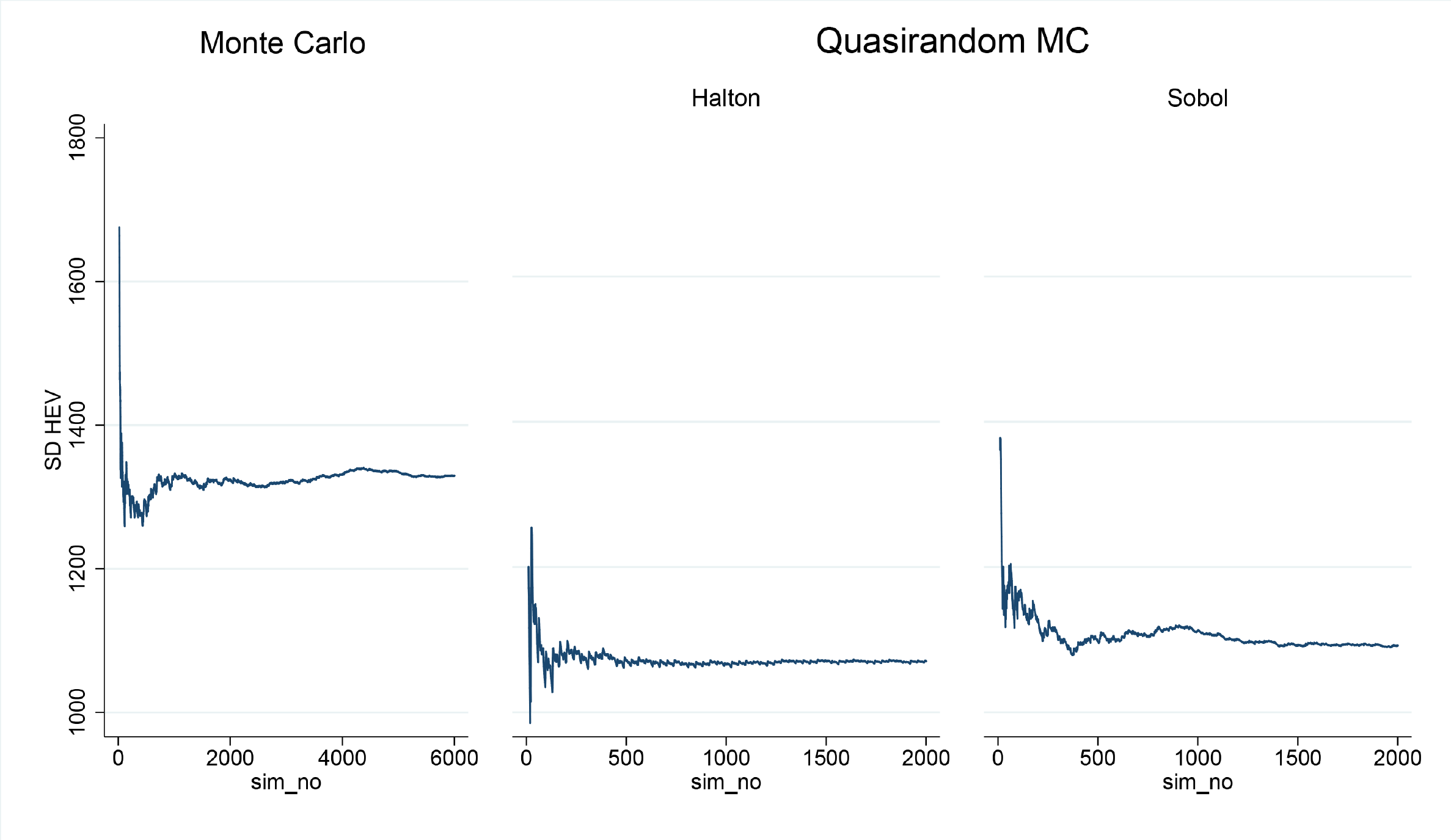}
	\caption{DIVA-C HEV Standard deviation changes for the triangular distribution using: random sampling; Halton; and Sobol sequences for Central Asia}
	\label{fig:CAS}
\end{figure}

In Central Asia (figure \ref{fig:CAM}) the mean is stable after approximately 4,000 simulations which are already 96 less than the QG method required as discussed above. In the Halton-based SSA results, we find that the mean converges to the rMC 10,000 runs mean (black line) after only 500 simulations and the confidence interval (CI) is similar at 2,000 simulations with the CI of 4,000 simulations in the rMC. Looking at the Sobol' QMC results, even though the mean converges at around 800 simulations, the Sobol'-based QMC is outperformed by the Halton QMC method, but is preferred from the rMC SSA. The same conclusion can be derived by looking at figure \ref{fig:CAS} of the results' standard deviation (SD). The SD is more stable in the Halton QMC method after 500 simulations compared to the 1,500 required in the Sobol' QMC. Clearly though, both quasi-random methods produce significantly lower SD estimates compared to the rMC results. 

\begin{figure}[H]
	\centering
	\includegraphics[scale=0.5]{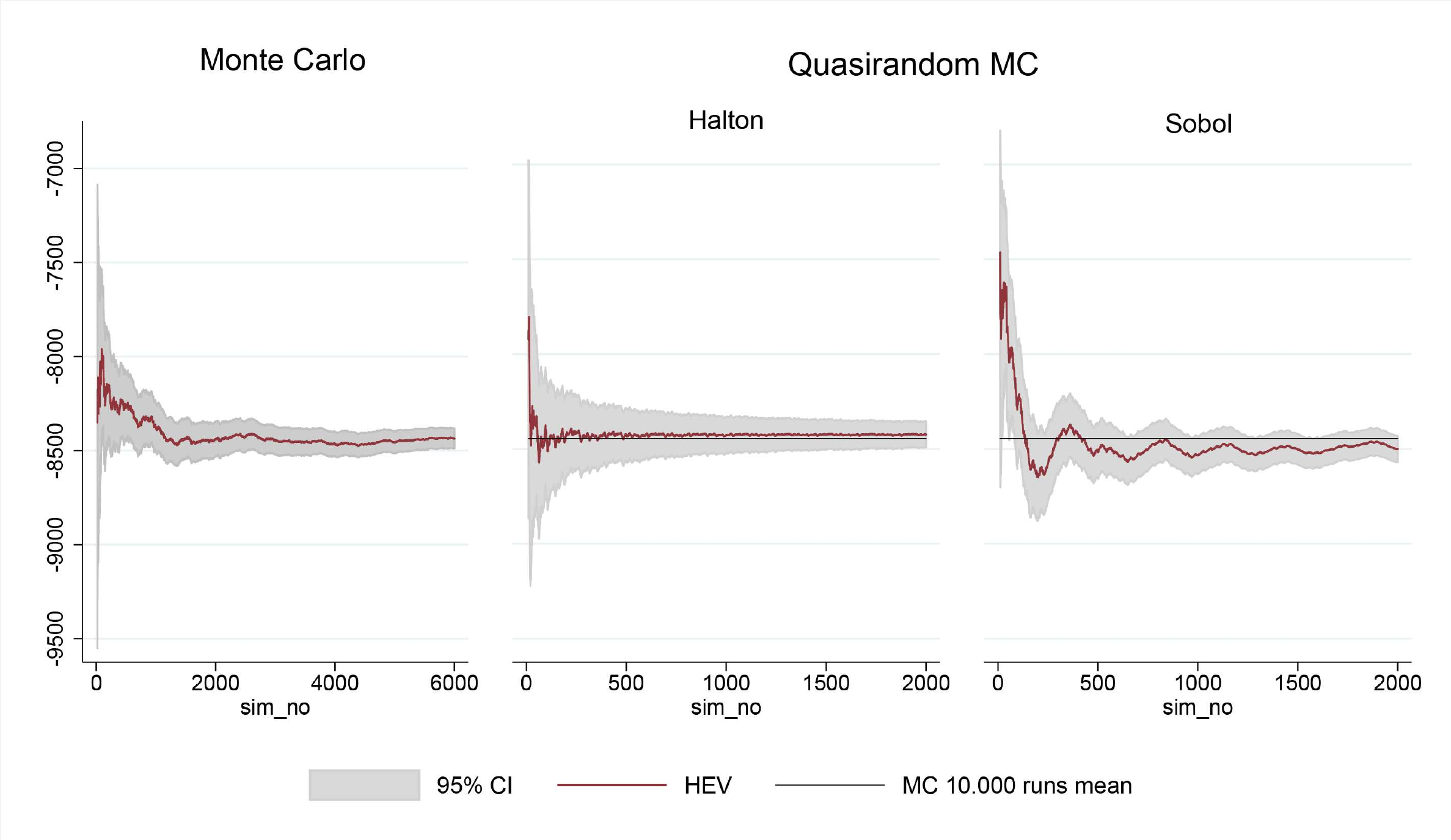}
	\caption{DIVA-C HEV mean convergence for the triangular distribution using: random sampling; Halton; and Sobol sequences for North America}
	\label{fig:NAM}
\end{figure}

\begin{figure}[H]
	\centering
	\includegraphics[scale=0.5]{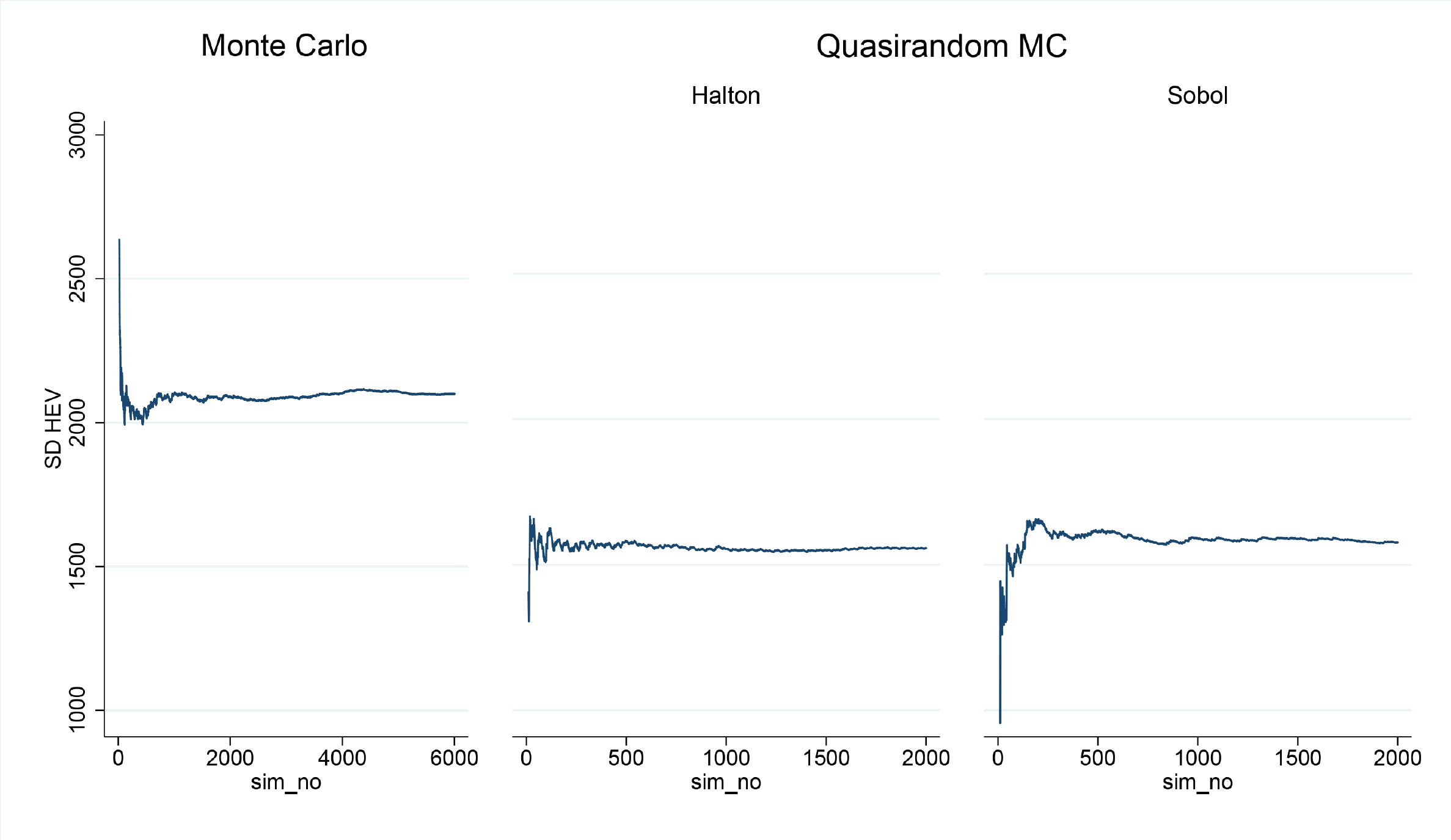}
	\caption{DIVA-C HEV Standard deviation changes for the triangular distribution using: random sampling; Halton; and Sobol sequences for North America}
	\label{fig:NAS}
\end{figure}

In North America (figure \ref{fig:NAM}), the superiority of the Halton-based method is more prominent. Clearly, the Halton-based results converge after 500 simulations, but the Sobol'-based results have not converged yet, before 2,000 simulations. Nevertheless, these results are superior to the 4,000 simulations required by the rMC method. The North American SD results (figure \ref{fig:NAS}) indicate that the SD reaches a plateau after 500 simulations in the Halton and 1,000 in the Sobol'-based results and that plateau is smaller than the rMC one.

\section{Conclusion}
In this paper, we discuss the use of Quasi-Monte Carlo methods based on the Halton and Sobol' sequences for systematic sensitivity analysis of CGE model. Our main interest is to provide evidence of efficiency gains with using QMC in SSA compared to rMC. 

Based on our empirical results, both Halton and Sobol' based QMC methods are more efficient than the rMC method, but the Halton-based results are preferred. This is based mainly on the convergence of the results. Our results indicate that there is a clear computational-time advantage in the use of low-discrepancy sequence compared to pseudo-random numbers and this method should be considered when conducting a SSA of a CGE model. In this paper we have only used two quasi-random sequences, but it would be interesting in the future to compare these results with other quasi-random sequences such as: the Faure sequence; the Hammersley set; and the Niederreiter sequence.

\section{Figures}

\end{document}